%% file: main.tex
\documentclass[sigconf]{acmart}
\AtBeginDocument{%
  \providecommand\BibTeX{{%
    \normalfont B\kern-0.5em{\scshape i\kern-0.25em b}\kern-0.8em\TeX}}}

\copyrightyear{2024}
\acmYear{2024}
\setcopyright{rightsretained}
\acmConference[CHI '24]{Proceedings of the CHI Conference on Human Factors in Computing Systems}{May 11--16, 2024}{Honolulu, HI, USA}
\acmBooktitle{Proceedings of the CHI Conference on Human Factors in Computing Systems (CHI '24), May 11--16, 2024, Honolulu, HI, USA}
\acmDOI{10.1145/3613904.3642487}
\acmISBN{979-8-4007-0330-0/24/05}

\acmConference[CHI '24]{Make sure to enter the correct
  conference title from your rights confirmation emai}{May 11--16,
  2024}{Honolulu, HI}
\acmBooktitle{CHI '24: ACM Conference on Human Factors in Computing Systems,
 May 11 -- 16, 2024, Honolulu, HI, USA} 
\acmISBN{979-8-4007-0330-0/24/05}

\begin{document}

\title[ClassInSight: Designing Conversation Support Tools to Visualize Classroom Discussion]{ClassInSight: Designing Conversation Support Tools to Visualize Classroom Discussion for Personalized Teacher Professional Development}



\author{Tricia J. Ngoon}
\email{tngoon@andrew.cmu.edu}
\affiliation{%
  \institution{Human-Computer Interaction Institute, Carnegie Mellon University}
  \country{USA}
}

\author{S Sushil}
\email{sushils@ucsd.edu}
\affiliation{%
  \institution{Education Studies, University of California, San Diego}
  \country{USA}
}

\author{Angela Stewart}
\email{angelas@pitt.edu}
\affiliation{%
  \institution{Department of Informatics and Networked Systems,  University of Pittsburgh}
  \country{USA}
}

\author{Ung-Sang Lee}
\email{ung-sang.lee@unlv.edu}
\affiliation{%
  \institution{Educational Psychology, Leadership, and Higher Education, University of Nevada, Las Vegas}
  \country{USA}
}

\author{Saranya Venkatraman}
\email{saranyav@psu.edu}
\affiliation{%
  \institution{College of Information Sciences and Technology, Pennsylvania State University}
  \country{USA}
}

\author{Neil Thawani}
\email{neilthawani@gmail.com}
\affiliation{%
  \institution{Human-Computer Interaction Institute, Carnegie Mellon University}
  \country{USA}
}

\author{Prasenjit Mitra}
\email{pum10@psu.edu}
\affiliation{%
  \institution{College of Information Sciences and Technology, Pennsylvania State University}
  \country{USA}
}

\author{Sherice Clarke}
\email{snclarke@ucsd.edu}
\affiliation{%
  \institution{Education Studies, University of California, San Diego}
  \country{USA}
}

\author{John Zimmerman}
\email{johnz@andrew.cmu.edu}
\affiliation{%
  \institution{Human-Computer Interaction Institute, Carnegie Mellon University}
  \country{USA}
}

\author{Amy Ogan}
\email{aeo@andrew.cmu.edu}
\affiliation{%
  \institution{Human-Computer Interaction Institute, Carnegie Mellon University}
  \country{USA}
}

\renewcommand{\shortauthors}{Ngoon et al.}

\begin{abstract}
Teaching is one of many professions for which personalized feedback and reflection can help improve dialogue and discussion between the professional and those they serve. However, professional development (PD) is often impersonal as human observation is labor-intensive. Data-driven PD tools in teaching are of growing interest, but open questions about how professionals engage with their data in practice remain. In this paper, we present ClassInSight, a tool that visualizes three levels of teachers’ discussion data and structures reflection. Through 22 reflection sessions and interviews with 5 high school science teachers, we found themes related to dissonance, contextualization, and sustainability in how teachers engaged with their data in the tool and in how their professional vision, the use of professional expertise to interpret events, shifted over time. We discuss guidelines for these conversational support tools to support personalized PD in professions beyond teaching where conversation and interaction are important.
\end{abstract}

\ccsdesc[500]{Human-centered computing~User studies}

\keywords{teachers, classroom discourse, teacher professional development, conversation support, reflection, data visualizations}


\maketitle

\input{sections/1_Intro}
\input{sections/2_RelatedWork}

\input{sections/3_Context}

\input{sections/4_CIS}

\input{sections/5_Method}
\input{sections/6_Findings}
\input{sections/7_Discussion}


\bibliographystyle{ACM-Reference-Format}
\bibliography{reference}

\appendix

\end{document}

%% file: sections/1_Intro.tex
\section{Introduction}
The classroom is a constantly changing environment. Both course content and standards for best practices in teaching constantly evolve. Teachers must adapt to the needs of individual students while also adapting to the group dynamics that make every class unique. Teachers need to plan lessons, reflect on the effectiveness of their practices, and develop strategies to engage students. Feedback is critical for understanding what is working, and finding opportunities for positive change \cite{d2010expert,nicol2006formative,shute2008focus}. Feedback can help teachers develop reflective practices toward improved student learning and greater equitable participation. Unfortunately, continuous feedback for teachers can be hard to come by \cite{jayaram2012breaking,archer2016better}. 

Class observations in which a professional with expertise in teacher training observes one or two class periods can provide personalized, in-depth feedback for teachers about their teaching practices \cite{archer2016better}. A review of studies found that such immediate feedback was most effective for specific and corrective changes to teaching behaviors \cite{scheeler2004providing}. However, observations often focus on performance evaluations of the teacher as an employee rather than personal growth \cite{stigler201824}. In addition, more subtle teaching moves, such as conversational dynamics during class discussions are difficult for a single observer to make sense of in real-time. While expert teaching observations are more frequent for pre-service teachers, they are less common for in-service teachers who would also benefit from continuous feedback. Video recordings offer an alternative, offering a more accurate recollection of events and rich opportunities for reflection \cite{dawson1975effect,blomberg2013five,borko2008video,sturmer2013changes}, but they require great effort. This includes setting up and testing recording equipment and effort to curate actions worthy of reflection from recordings \cite{blomberg2013five,derry2010conducting,jacobs2007video,goldman2014video,sherin2008professional}. More generic forms of professional development (PD) such as workshops and seminars can scale and provide background on changing content and teaching standards, but they cannot provide the personalized and persistent guidance necessary for effective change \cite{jayaram2012breaking,popova2022teacher,ball1999developing,sherin2008professional}. In addition, most teacher PD sessions are singular instances that do not provide opportunities for progress follow-ups on new techniques and approaches. Finally, PD largely focuses on standards or content rather than individual goals, strategies, and strengths \cite{jayaram2012breaking,jacob2015mirage}. 

Learning scientists have recently investigated technology for observing teaching practices as a novel type of personalized PD, particularly for discourse pedagogy. Class discussion can give voice to students' reasoning and provide teachers with a greater understanding of student cognition and learning process \cite{clarke_dialogic_instruction,gillies_productive_academic_talk,asterhan2015socializing,resnick2010well}. Prior work has found that even low cost recording equipment could accurately capture and model conversational dynamics in the classroom \cite{d2015multimodal,jensen2020toward}. Several current systems, both research and commercial, use class audio recordings to provide teachers with automated feedback about their discussion practices \cite{jensen2020toward,jensen2021deep,demszky2023can,gomoll2021zooming,ford2021student,chen2015classroom,d2015multimodal}\footnote{https://teachfx.com}.  
Open questions remain about the design of such tools and their impact in practice, which are relevant to the HCI and educational technology communities for bridging research and practice. For our work, we build upon ClassInSight as a tool that visualizes classroom discussion data for personalized teacher PD and conversation support \cite{gomoll2021zooming}. Our approach to designing ClassInSight incorporates data visualizations of discourse and collaborative reflection with a PD researcher to answer the following research questions: 

\begin{itemize}
\item How did teachers use features of ClassInSight in reflections over time?
\item What are the factors and barriers of adoption of discourse visualization tools?
\end{itemize}

As part of a collaborative design-based research (DBR) \cite{design2003design} project among three research institutions, we situate ClassInSight within middle and high school science teaching and guided reflection sessions in which teachers discuss their class discussion data with a PD researcher. This tool is an expansion of Gomoll et al \cite{gomoll2021zooming} that visualizes classroom discussion data in three different levels: the Talk Ratio, Turn-Taking, and Transcript (Figure \ref{fig:interface}). In addition to these visualizations, teachers follow a schema to structure their noticings and reflections. As part of a longitudinal deployment over 3 academic years, 5 middle and high school science teachers from a large city in the United States participated in 22 reflection sessions during which they engaged with their discussion data in the tool. At the end of the deployment, we conducted interviews with teachers to understand their experiences in using the prototype. From their interactions with the tool in reflection sessions and interviews, we found themes related to quantification, contextualization, shifting professional vision, and adoptability. We extend previous results \cite{gomoll2021zooming,tripathi2022designing} by showing how interactions with data within ClassInSight impact reflections over time.  

We make the following contributions. First, we provide an analysis of how the design of a data-driven discussion analysis tool, ClassInSight, impacts professional learning and reflection in teaching. Second, we contribute design implications that influence the adoption of conversational support tools in professions that often also lack frequent personalized feedback on professional interactions (\textit{e.g.} clinicians, vets, mentors, therapists, trainers, police, advisors, lawyers, consultants). Our findings from the deployment of ClassInSight provide lessons learned both within and beyond teaching.


%% file: sections/2_RelatedWork.tex
\section{Related Work}

\subsection{Teacher Professional Development (PD) Practices \& Professional Vision}
How teachers notice and interpret events of their classroom is important in adapting to the various unexpected scenarios that can occur. As much as teachers plan their lessons, they also need to be able to respond to students and encounters on-the-fly \cite{sherin2008professional,wallach2005hearing}. Professional vision, defined as the use of professional expertise and knowledge to interpret events related to professional interactions \cite{goodwin2015professional}, encompasses this knowledge. Through development of professional vision, teachers are able to describe, explain, and predict classroom scenarios \cite{seidel2014modeling}. This concept is related to noticing, the ability to discern and attend to the consequential features of instruction during reflection \cite{van2021expanding}. Professional development (PD) can help teachers in developing professional vision. It is the continuous education and training of a professional worker to update skills and expertise  \cite{cervero2001continuing, dall2006unveiling, grant1992obsolescence, mizell2010professional,easton2008professional,o2003unlocking}. Recent literature has shifted to framing PD as professional learning, which emphasizes the view that professionals undergoing PD are continuous learners themselves \cite{mann2022thinking,stewart2014transforming,webster2009reframing,rppl}. Koellner and Jacobs \cite{koellner2015distinguishing} present a model of PD that is adaptive to individual teachers as they progress in their professional learning. 


Feedback is an critical component of PD and developing professional vision. Continuous, formative feedback can help professionals in understanding their current performance and knowing how they can best improve. However, personalized feedback is rare because most PD consists of single-session workshops or seminars that are not tailored to the professional's context \cite{jayaram2012breaking,popova2022teacher,ball1999developing}. Consultations can provide more personalized PD, but are often one-offs and are infrequent after initial professional training, as for pre-service teachers \cite{watling2013beyond,jayaram2012breaking}. Additionally, quantifying certain measures that are not performance-based, such as dialogue, emotions, or non-verbal behaviors, can be challenging. 

\subsection{The Importance of Discussion in the Classroom}
Classroom discussions where both students and teachers are actively engaged in learning can help students explicate their reasoning \cite{asterhan2015socializing,clarke_dialogic_instruction,gillies_productive_academic_talk,resnick2010well}. It gives students a voice in their learning, helps them reason more deeply, and expand their evaluations of arguments \cite{clarke_dialogic_instruction,gillies_productive_academic_talk,asterhan2015socializing,resnick2010well,michaels2008deliberative,resnick2018next}. Teachers can uptake student ideas through following up with questions or further elaboration to promote further student engagement \cite{collins1982discourse}. Productive discussion can lead to improved curriculum mastery, reasoning, and educationally relevant attitudes in math, science, and literacy classes \cite{howe2019teacher}. Learning scientists have explored frameworks for how teachers can use discourse to engage with their students. For example, Resnick et al \cite{resnick2010well} explain the concept of Accountable Talk to socialize discussion into a community of practice that normalizes grounded argumentation and prompts students for further responses or counterarguments. The framework of academically productive talk outlines how teachers' talk moves such as asking students to restate ideas or asking students to elaborate on reasoning can socialize and encourage discussion \cite{resnick2010well,michaels2008deliberative}. Michaels and O'Connor \cite{michaels2015conceptualizing} conceptualized productive talk within four goals for students: sharing their thoughts, listening to others, deepening reasoning, and engaging with others' ideas. These discourse frameworks can also provide analytical tools for understanding how talk is used in a class. The Scheme for Educational Discourse Analysis (SEDA) \cite{hennessy2016developing} and more recently, the Teacher Scheme for Educational Discourse Analysis (T-SEDA) \cite{vrikki2019teacher} are coding schemes that highlight sequences of dialogue in classrooms. Talk moves include inviting, guiding, building, connecting, evaluating, and reflecting on ideas to promote classroom discussion \cite{vrikki2019teacher,hennessy2016developing}. In this paper, we adopt the T-SEDA coding scheme for our dialogue coding and visualizations.

Creating a learning environment conducive to productive discussion takes deliberate and conscious effort potentially over months \cite{michaels2008deliberative}. Teachers may not know how to engage students in deep reasoning and fear that little to no students will respond \cite{michaels2015conceptualizing}. Teachers may also find argumentation uncomfortable in general and have difficulty guiding students in respectful evaluation and argumentation \cite{resnick2018next}. Novice teachers in particular may not have experience in noticing moments where they can strategically guide students' discussion \cite{clarke2016uncovering}. As a result, discussion that promotes deep reasoning and evaluation is far less common than discussion based on close-ended questions and short responses \cite{mcneil_discourse_urban_classrooms,resnick2018next}. This provides opportunities for PD to support teachers' use of discussion in their teaching \cite{hennessy2019teacher,tripathi2022designing,hennessy_PD_dialogic_teaching}. For instance, Calcagni et al \cite{calcagni2023developing} reported on a 15-month study with 20 institutions incorporating T-SEDA and found that teachers felt more agentic in changing their dialogic practices. Video feedback to contextualize and replay specific moments of discourse has been shown to improve productive talk use and student learning outcomes \cite{jacobs2007video,borko2008video,chen2020efficacy,sherin2008professional}. From these promising results, technology to support discourse PD is a prominent area within learning science and HCI research.

\subsection{Conversational Support Tools for PD}
Both learning science and HCI communities have explored technology to scale conversational support towards developing teachers' professional vision through discussion. This involves recording, either through video, audio, or both, class events for later reflection. Prior work has examined how well automated models can accurately identify teacher and student discourse variables from recordings collected from low cost, easy to use recording equipment \cite{d2015multimodal,jensen2020toward,jensen2021deep,schlotterbeck2021classroom}. For example, Jensen et al \cite{jensen2020toward} presented a model that performed within the range of a human observer in identifying discourse variables such as instructional talk, elaborated evaluation, and authentic questions. Cook et al \cite{cook2018open} also found that combining models can lead to improved performance in detecting questions from classroom discussion.

In addition to the breadth of work in modeling of classroom discourse, several tools visually represent this data for feedback to teachers. Chen et al \cite{chen2014analytic,chen2015classroom} explored using Classroom Discourse Analyzer (CDA) to show the distribution of teacher and student talk through visual representations of turn-taking patterns within a class session. A randomized controlled trial with CDA showed that teachers increased productive talk moves and more easily navigated classroom video recordings CDA's interface \cite{chen2020efficacy}. The TalkMoves application \cite{jacobs2022promoting} transcribes and visualizes classroom talk in a dashboard according to six types of Accountable Talk instructional moves (Keeping everyone together, Getting students to relate another's ideas, Restating, Pressing for accuracy, Revoicing, Pressing for reasoning) to give immediate, automated feedback to teachers. Researchers found that teachers that utilized the TalkMoves application improved their talk moves in K-12 math courses \cite{scornavacco2022automated}. A randomized controlled trial with the M-Powering Teachers tool, which displays percentage of student and teacher talk alongside a transcript of the discussion, demonstrated an increase in teacher uptake of student ideas through acknowledgements, questions, and revoicing by 13\% \cite{demszky2023can}. Gomoll et al \cite{gomoll2021zooming} explored how different granularities of visual representations of their talk moves can help teachers shift create nuanced constructions of their professional vision related to classroom discussions. In the commercial space, TeachFX is a mobile application in which teachers can self-record their classes and view how much they talk and engage students in discussion. A pilot study with TeachFX showed a 45\% increase in student talk, particularly in Black and Brown students \cite{ford2021student}. There is promise in these tools for improving teacher discourse in the classroom. How the design of these systems and data visualizations facilitate teacher reflection as well as how teachers incorporate these systems in practice are open areas of work, which we address in this paper. 



%% file: sections/3_Context.tex
\section{Research Context}

\begin{table*}[]
\begin{tabular}{r|ccccc|}
\multicolumn{1}{c|}{\textbf{\begin{tabular}[c]{@{}c@{}}Participant \\ Pseudonym\end{tabular}}} & \textbf{Gender} & \textbf{\begin{tabular}[c]{@{}c@{}}Teaching\\ Experience\end{tabular}} & \textbf{\begin{tabular}[c]{@{}c@{}}Grades \\ Taught\end{tabular}} & \textbf{Current Subject Taught} & \textbf{\begin{tabular}[c]{@{}c@{}}\# of Reflection \\ Sessions\end{tabular}} \\ \hline
Jeff & M & 15 & 9-12th & Engineering & 6 \\
Tom & M & 15 & 9-12th & Earth Science, Chemistry & 6 \\
Sheila & F & 14 & 7th & Science & 2 \\
Bonnie & F & 18 & 6th & Science & 5 \\
Kate & F & 17 & 9-12th & \begin{tabular}[c]{@{}c@{}}Earth Science, Chemistry, \\ Environmental Science\end{tabular} & 4
\end{tabular}
\caption{Teacher demographics by pseudonym, gender, teaching experience, grades taught, current subject being taught, and the number of reflection sessions teachers participated in.}
\label{tab:demographics}
\end{table*}

This paper presents the latest cycle of a design-based research (DBR) project with collaborations between three research institutions. As part of a research-practice partnership, researchers and teachers collaborated in a long-term partnership to address problems of practice rather than problems of theory \cite{coburn2016research}. Our interdisciplinary research team consists of faculty, postdoctoral researchers, graduate students, and designers with expertise in learning science, human-computer interaction, design, natural language processing, and software development. We met weekly to discuss design directions and decisions. DBR is complementary to human-centered design as it involves iterative cycles of designing interventions and testing these interventions in educational contexts \cite{design2003design}. DBR addresses learning in authentic educational contexts beyond narrow measures of learning \cite{collins1992toward}. Our goal with this DBR project was to design a tool that helps teachers facilitate discussions in science classrooms. As developing discourse practice takes effort and time \cite{michaels2008deliberative}, we chose a longitudinal approach to examine teachers' reflections and learning in-depth over 3 academic years. This approach is rooted in field trials and longitudinal studies to understand users' experience in HCI \cite{kjaerup2021longitudinal,forlizzi2004understanding}. This study was approved by the Institutional Review Board (IRB).   

\subsection{Teacher Participants}
\label{sec:participants}
From the 2019-2022 academic years, 5 in-service middle and high school science teachers (3 identified as female, 2 identified as male) from 5 schools in the same public school district in a large city in the southwestern United States participated in interviews, co-design sessions, reflection sessions, and prototype testing with our team. All teachers had at least 10 years of teaching experience (average 15.8 years) and taught various science subjects at the middle and high school levels. Table \ref{tab:demographics} shows demographic information for our participants and the pseudonyms we will use throughout the paper for each teacher. All teachers and the parents of students in their class consented to participating in our research. From these interactions, we developed an early version of ClassInSight and deployed it with teachers in collaborative reflection sessions \cite{gomoll2021zooming}. This paper discusses the latest design of our ClassInSight prototype and findings from reflection sessions and final interviews with teachers. As part of this longitudinal project took place during the COVID-19 pandemic, a large part of our data collection in classrooms and with teachers took place over Zoom videoconferencing\footnote{https://zoom.us}.


%% file: sections/4_CIS.tex
\section{ClassInSight: A Dialogue Visualization Tool for Personalized PD}
In this section, we describe the main features of ClassInSight, including three data visualizations of classroom discussion and a structured schema to guide reflections. We also describe each step of our workflow in data collection, data coding, and data visualization in the interface.

\subsection{Discussion Data Collection}
D'Mello et al \cite{d2015multimodal} defined a set of constraints for recording audio in classrooms including the ease of use by teachers, affordability of equipment, non-intrusiveness, and quality of recordings for utility. Following these constraints, we obtained a pressure zone microphone (PZM), a lapel mic, an audio mixer, and a laptop with Audacity to record the audio through the mixer pre-COVID. A researcher scheduled a visit to the participant teacher's classroom to set up the equipment and record classroom discussions. During virtual instruction during COVID, a researcher recorded online class discussions virtually. When classroom instruction resumed in-person, the research team sent recording equipment to teachers and provided instructions on how teachers could record discussions on their own. A researcher remained available over phone for real-time technical support. In total, teachers recorded 62 class sessions. All audio recordings were transcribed with a third-party transcription service\footnote{https://rev.com}.


\subsection{Dialogue Categorization \& Coding}
\label{sec:coding}
Multiple studies have shown that teachers' intentional use of dialogue to guide students' sensemaking is beneficial for their long-term retention, domain transfer, and reasoning development \cite{clarke_dialogic_instruction,littleton2010educational,asterhan2015socializing}. With this premise, we intended to develop teachers' noticing and reflection of productive elements of their dialogic practice in the classroom. In our previous iterations, we conceptualized discussion around \cite{clarke2016student}, but found that developing a shared meaning of these talk codes between researchers and teachers constrained the collaboration in reflection sessions \cite{tripathi2022designing}. Thus, in the current version of ClassInSight, we adapted the works of \cite{wells2006dialogue} and \cite{hennessy2016developing} to use a dialogue coding scheme according to the following codes: \textit{Build on Ideas, Connect, Evaluation, Express or Invite Ideas, Guide Direction of Dialogue, Invite Elaboration or Reasoning}, and \textit{Make Reasoning Explicit}. Miscellaneous talk that did not fall within these categories was labeled as \textit{Other Classroom Talk} and \textit{Other Outside Talk}. A subgroup from the research team worked together to achieve inter-coder reliability to in a complex and long process that spanned two academic years to achieve a Cohen's Kappa of .70 or above between expert and novice coders which was considered \textit{very good agreement} \cite{clarke2023developing}.

\begin{figure*}[t]
\centering
   \includegraphics[width=\textwidth]{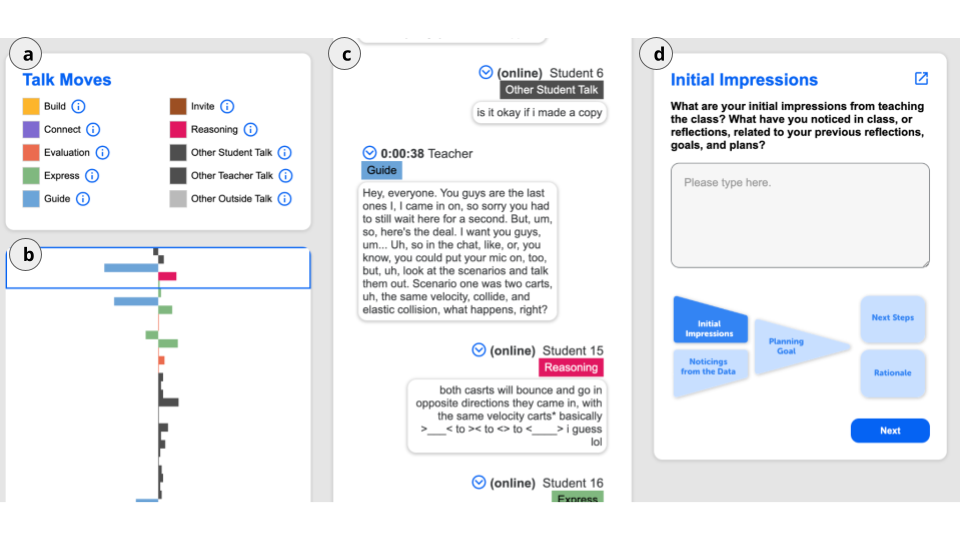}
   \caption{An overview of the ClassInSight interface: a) The legend with talk codes, color coding, and description, b) The miniview showing the Turn-Taking visualization as a tree visualization, c) The Transcript visualization shows the full transcript of dialogue, d) the schema that guides reflection during collaborative reflection sessions.}
\Description{A web application interface for our tool. a) In the top left corner is a legend containing the description of talk moves and their associated color codes. b) On the bottom left is a miniview of the Turn-Taking visualization. c) In the middle, the Transcript visualization shows the transcript of the class. Each piece of dialogue contains the associated talk codes. d) On the right of the interface is the structured reflection schema where teachers and PD researchers can take notes related to their noticings.}
\label{fig:interface}
\end{figure*}

\subsection{ClassInSight Interface}
After categorizing classroom discussion transcripts, the data is visualized in the ClassInSight interface. Similar to our previous version of the ClassInSight \cite{gomoll2021zooming}, the interface features three levels of visualizations: the Talk Ratio visualization, the Turn-Taking visualization, and the Transcript visualization, each highlighting discussion at different granularities. This final version includes improved design and interaction from teacher feedback as well as a structured schema to guide teachers’ reflections. We describe each of these features in this section. ClassInSight is implemented as a web application in React.js.

\begin{figure*}[t]
\centering
   \includegraphics[width=\textwidth]{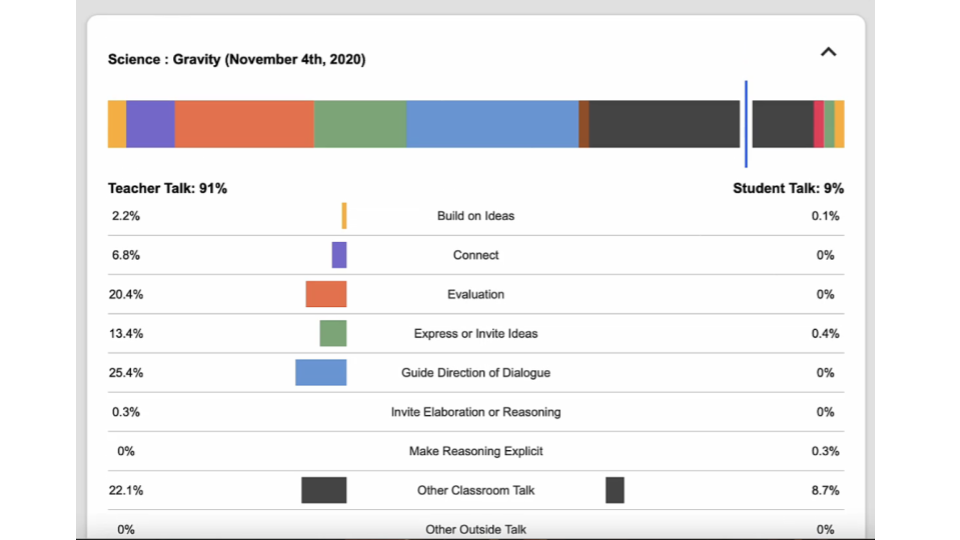}
   \caption{The Talk Ratio visualization, which shows the overall classroom discourse in terms of talk category. The vertical line separates the percentage of teacher talk versus student talk. Below the visualization is a breakdown of each talk move by percentage. Clicking on any talk move shows excerpts from the transcript if available.}
\Description{The Talk Ratio visualization shows a horizontal bar graph that contains breakdowns of the percentage of each talk move present. A vertical line in the bar graph delineates the percentage of teacher versus student talk (the image shows Teacher talk: 91\% and Student talk: 9\%). Below the graph are each of the talk moves. Clicking on the talk move shows excerpts of that talk move if available.}
\label{fig:talkratio}
\end{figure*}

\subsubsection{Talk Ratio Visualization}
The Talk Ratio visualization (Figure \ref{fig:talkratio}) gives a coarse-grained summary of discourse. It is on the first page that teachers and researchers see in their reflection sessions. The Talk Ratio visualization separates class discourse in terms of the percentage of teacher versus student talk. This is similar to existing tools that show proportion of talk in teacher PD \cite{demszky2023can} and other professions \cite{hirsch2018s}. Discourse is further broken down according to the coding scheme discussed in Section \ref{sec:coding} to show the percentage of different types of talk and color codes for each category. This provides a quantified overview of who is talking and what types of talk are occurring.

\begin{figure*}[t]
\centering
   \includegraphics[width=\textwidth]{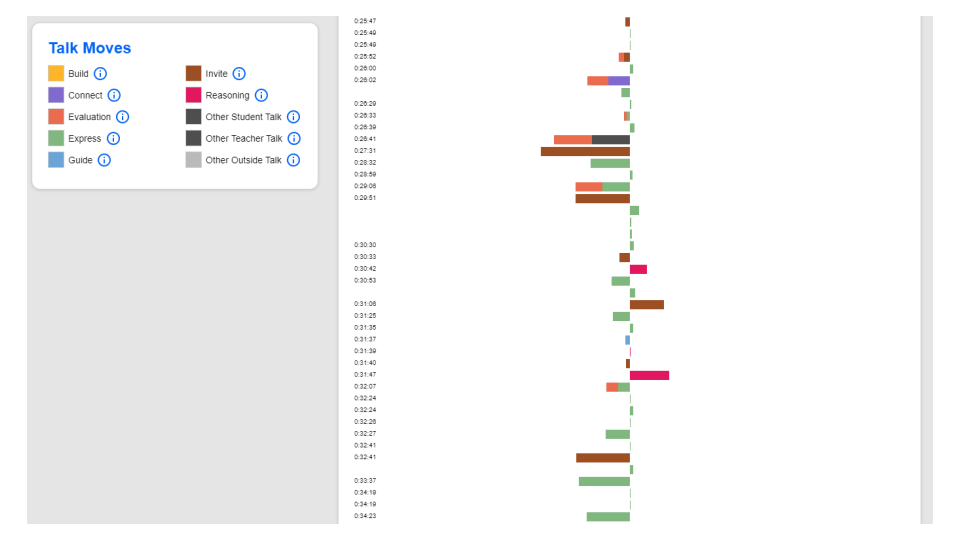}
   \caption{The Turn-Taking visualization shows the rhythm or cadence of the discussion with teacher talk on the left and student talk on the right. Each bar represents a piece of dialogue and is proportional with the length of dialogue. The colors of the bars represent the applicable talk to that specific piece of dialogue. A legend on the left displays the code labels and colors.}
\Description{The Turn-Taking visualization shows the legend of talk moves and associated color codes on the left. On the right is the tree visualization, which shows the cadence of conversation, with teacher talk on the left, and student talk on the right. Each piece of dialogue is represented as a bar that is proportional to its length. Each bar is color coded according to the talk move(s) associated with it.}
\label{fig:turntaking}
\end{figure*}

\subsubsection{Turn-Taking Visualization}
The Turn-Taking visualization (Figure \ref{fig:turntaking}) is a tree visualization that features teacher talk on the left and student talk on the right. It shows a timeline of the discussion and the cadence of the discussion. The bars represent the length of each piece of dialogue, which is defined by how long each individual spoke. Each bar displays the colors for all codes that apply to that piece of dialogue. The visualization is less quantified in terms of concrete numbers and more qualitative about the sequence of discussion between teachers and students. A legend (Figure \ref{fig:interface}a) in the upper left corner of the interface shows the color and labels for all codes as a visual reminder. An ``i'' icon beside each code provides a short definition that teachers can hover over to see.

\subsubsection{Transcript Visualization}
The Transcript visualization (Figure \ref{fig:interface}c) provides the greatest amount of detail, showing a transcribed reading of the class discussion and code labels above each dialogue piece. To help users maintain understanding of where each piece of dialogue occurs in the whole discussion, the miniview of the Turn-Taking visualization (Figure \ref{fig:interface}b) is visible on the bottom left, and a box outlines where in the Turn-Taking view each piece of dialogue takes place and adjusts based on scrolling in the transcript.

\subsubsection{Interaction between Visualizations}
Similar to the interactive visualizations in Chen et al \cite{chen2015classroom}, each visualization in our previous iterations was on its own page and disconnected from the context of the other visualizations. A significant change in the current version is that each visualization is interactively connected to each other, allowing teachers to make connections between the visualizations. For instance, in the Talk Ratio visualization, clicking on the part of the visualization pertaining to Evaluation shows a list of the excerpts related to the Evaluation code from the Transcript visualization. Clicking on an individual excerpt would navigate to the location of that excerpt within the Turn-Taking and Transcript visualizations. Similarly, clicking on a specific section within the Turn-Taking visualization would navigate to the location of that dialogue within the Transcript visualization. Following design recommendations from \cite{sharif2021understanding}, this interaction enables users to both holistically view their data in the Talk Ratio and drill-down on specific data points via the Turn-Taking and Transcript visualizations.

\subsubsection{Structured Reflection Schema}
Previously, researchers and teachers took notes in a separate location on their own during reflection sessions. To keep the context of reflection sessions and data in one place, we included an interactive schema (Figure \ref{fig:interface}d) to structure reflection. The schema is persistent across all visualizations on the right side of the interface so researchers and teachers can add notes at any point during their reflection. The schema provides text boxes for reflections regarding \textit{Initial Impressions, Noticings from Data, Planning Goal, Next Steps, }and \textit{Rationale}. These sections are based on the reflection and action cycles within the framework of personal informatics \cite{li2010stage} and in the action-reflection-planning framework in teacher PD \cite{gerritsen2018towards}. The sections of the schema were shaped into a triangle to provide a flow and structure for reflection during collaborative reflection sessions. Users can add notes within the schema text boxes according to the specific section.

%% file: sections/5_Method.tex
\section{Method}
Here we describe our reflection session and interview procedures with participants and our data analysis of reflection session transcripts and interviews. All reflection sessions and interviews took place on Zoom video conferencing and were transcribed using Rev.com. 

\subsection{Guided Reflection Sessions}
\label{sec:reflectionsessions}
The goal of reflection sessions was for teachers to reflect on classroom discussion visualizations in a collaborative setting with a PD researcher, identify changes to make in their practice based on these noticings, and plan appropriate next steps. These reflections took place at least once per academic year and were scheduled by email between one of two PD researchers and the teacher. All reflections were roughly 40 minutes to an hour long. Due to the COVID-19 pandemic and challenges in scheduling, reflection sessions took place at variable time frames, with the shortest amount of time between reflection sessions being 1 month and the longest being 16 months. After a class discussion was recorded and data was processed, coded, and visualized in the tool, teachers were given the option to review their data in preparation. During reflection sessions, the researcher first reviewed what was discussed in the last reflection session or discussed the goal of the reflection. The researcher then guided the teacher in navigating the visualizations in the tool starting with the Talk Ratio visualization. Following the structured reflection schema (Figure \ref{fig:interface}d), the researcher and teacher first discussed teachers' noticings in the visualizations, how they made sense of it, and what goals to set based on their data. All teachers participated in at least 2 reflection sessions, with the maximum number of reflection sessions being 6 (see Table \ref{tab:demographics}). 

\subsection{Participant Interviews}
At the end of the deployment, we contacted teachers to participate in interviews regarding their experiences engaging with our prototype. We interviewed 4 of the 5 teachers as one teacher (Bonnie) was unable to participate due to scheduling conflicts. Through hour-long semi-structured interviews, we asked questions about how teachers utilized the visualizations, what they noticed in their data visualizations, and comparisons to existing PD practices. For their time and participation, teachers received a \$40 USD Amazon gift card. 

\subsection{Data Analysis}
To address our research questions, we first analyzed data using thematic analysis \cite{braun2006using}. The first and second authors first read through post-interview and reflection session transcripts. This amounted to roughly 25 hours of data total. As our data was longitudinal, we considered themes both for each individual teacher and across time \cite{thomson2003hindsight}. Our approach was inductive, developing codes as patterns within the data emerged. We then used affinity diagramming to group 353 quotes from reflection sessions and 72 quotes from post-interviews together based on emerging themes. Affinity diagramming is a common qualitative analysis technique that involves iteratively grouping relevant data into higher-level themes or affinities \cite{holtzblatt1997contextual}. The first and second authors then discussed these themes together and iterated on them until a consensus was reached and shared with the rest of the research team. 

%% file: sections/6_Findings.tex
\section{Findings}
We connect main themes of \textit{quantification, context, shifting professional vision} and \textit{adoptability} to our research questions of how teachers used features of the tool in their reflections over time (Sections 6.1, 6.2, and 6.3) and the factors and barriers to adoption of discourse analysis tools (Section 6.4). Each quote from a reflection session is labeled with (R) and the number of the reflection session (\textit{i.e.} (R2) would denote a quote from a teacher's second reflection session). 

\subsection{Quantification of Discussion Data in the Talk Ratio Visualization}
\label{sec:quant}

The Talk Ratio gave a quantified summary of teacher and student talk and the types of talk that occurred. Teachers noted this in their post-interviews. In particular, 2 teachers mentioned how quantification of discussion gave them insights into their facilitation practices. As Kate stated, ``\textit{I do think it’s good to see the categorization and the amount of time or the relative amount of time...to see how much time is spent on the different types of communication, and I do like the bar graph indicating when there's a lot more or a lot less.}'' Sheila thought the Talk Ratio visualization helped her see her class from an external perspective, ``\textit{The whole point of being able to see and being able to have like that fishbowl of the dynamics in my classroom is really helpful.}'' As the first visualization teachers saw, the Talk Ratio gave a coarse overview of the discussion from which they could form initial impressions. 

\subsubsection{Dissonance from the Talk Ratio Quantification}
Seeing the Talk Ratio often led to initial surprise or dissonance throughout all reflection sessions for teachers. For instance, Jeff mentioned, ``\textit{It's odd that this one was 82\% [teacher talk] to 18\% [student talk], because...I thought this one might be more [like] 60\% to 30\%. So I'm a little surprised that it went more the other way}'' (RS4). This dissonance continued into later reflections as well, ``\textit{I was still surprised at how much of the talking came [from] my side because I thought that it was still much more the students, but it appears that it was still majority me according to this}'' (RS6). Similarly, Bonnie experienced dissonance from the Talk Ratio, ``\textit{So, this is a little surprising to me to have less student talk this time, and I, in my head, I'm like, `Why?'}'' (RS4). This dissonance likely occurred because the data differed from teachers' expectations about their talk practices. As Kate noted, ``\textit{It's always shocking to me how much time I spend talking}'' (RS2). 

In post-interviews, teachers mentioned how the dissonance experienced from seeing the Talk Ratio provided a different perspective to their discussion. Jeff noted, ``\textit{The thing that I found most striking is I'm always talking more than I think I am so to see the graph of it is really helpful to let me know how much I need to get the students talking more.}'' He further elaborated, ``\textit{What seems like a normal amount seems like a lot...But it turns out that because a lot was only 15\% before that, they were only talking 19\% now, and it seems like they're talking a lot more than usual. Then [I’ll see] the graph and say, `oh wow they were only talking 19\%!}''' Kate also mentioned this dissonance in terms of talk codes in the Talk Ratio, ``\textit{There's a personal perception and then there's actually seeing the data, and so for me to see how much time I really did spend on Building and Connecting versus Evaluating or other talks helped a lot.}'' The Talk Ratio was a way for teachers to see how much their expectations were accurate to the discussion data.

\subsubsection{Learning from Dissonance}
According to the theory of cognitive dissonance, dissonance leads to people's need to resolve it \cite{cancino2020general}. In our study, teachers resolved their dissonance by adding to their beliefs about their talk and setting new discussion goals. Jeff noted, ``\textit{I think in class, hopefully [the Talk Ratio] will get closer to 50/50...That might still be doable in the near future...But I'm still working on pausing more, giving students time to think and respond}'' (RS3). 2 teachers talked about their discussion goals in terms of ``moving the line'' in the Talk Ratio visualization. Jeff said, ``\textit{I still wanna move this line left and get more students talking}'' (RS4). Sheila also mentioned, ``\textit{Toward the end of the year, that line shifts where it's mostly, that's where you want it to be, is that it's mostly student}'' (RS2). Dissonance led teachers to think about whether and why their data confirmed or contradicted their expectations. Bonnie discussed, ``\textit{I like that [the Talk Ratio] breaks it down by percentages so I can see,...is this where I want it to be? And it helps me start thinking about are these the results that I expected or are they different from the results I expected?}'' (RS2). Tom stated his expectations in terms of concrete percentages, ``\textit{I would think that Building on Ideas should be an important part of a lesson...Build on Ideas, 4\% doesn't look very substantial}'' (RS3). In post-interviews, 2 teachers mentioned using the Talk Ratio to ask questions about their discussion behaviors. Tom stated, ``\textit{I think it's a positive thing when students are talking...What are the things we can do to get students to participate...in the class?}'' Sheila also said, ``\textit{Am I giving them an appropriate amount of time and opportunity in order to answer questions? Am I actually asking questions that require them to answer using more than one word?}''  For these teachers, dissonance led to a desire for greater understanding about their discussion data. 

\subsection{Contextualization: Recalling and Understanding Dialogue Data in the Turn-Taking and Transcript Visualizations}
From the quantification of talk in the Talk Ratio, teachers then used the Turn-Taking and Transcript visualizations to further understand and contextualize what occurred through recalling moments and gaining an understanding of discussion dynamics.

\subsubsection{Recalling Classroom Events in the Transcript}
To understand what happened in a class, teachers used the Transcript visualization to recall specific moments. Jeff referred to specific lines in the Transcript,  ``\textit{And then...line 290, [I am] kind of just reassuring them again like, `Hey, like nothing to be ashamed about.'''} (RS2). Kate also used the Transcript to recall and explain events, ``\textit{So yeah, here's where we're getting into the article that we read, and yeah, so here's where they're actually starting to classify}'' (RS4). Tom added that he sometimes found surprise in the Transcript, ``\textit{It's always weird when I see what I actually say in class. I find it so strange because it's totally different than my impression of what I'm saying}'' (RS3). Recollection of discussion provided teachers with context into what was said and strategies used. As Bonnie noted the value of this recollection, ``\textit{The value that I see in...having a transcript...[is] to be able to look at the actual conversations...You still have to look at the actual content.}'' (RS4).

\subsubsection{Understanding Classroom Dynamics in the Turn-Taking and Transcript}
Where the Transcript helped teachers recall moments in class, the Turn-Taking guided teachers towards where to look. As Kate noted, ``\textit{I will say this turn-taking is kind of also eye-opening in terms of making a more visual representation of the percentages from the Talk Ratio}'' (RS2). She noticed, ``\textit{There's long stretches where I'm the only one talking a lot}'' (RS2). Tom identified ``chunks'' of teacher dialogue, ``\textit{To start off, I seem to be saying larger chunks, and then there's a period of time where [the chunks are] quite short and then they get longer again, towards the end of the class}'' (RS4). He noted how these chunks led him to examine the Transcript, ``\textit{So [I'm] looking at the colors and...at what I'm saying, and then what the students are saying, and trying to relate those two, which I think is exactly what you would wanna do with this kind of information}'' (RS4). Jeff also noticed chunks in his discussion data, ``\textit{So I’m seeing like big blocks of me [talking] still. But it seems like there’s some good back and forth}'' (RS2). Teachers used the Turn-Taking and Transcript to answer questions about their data. For example, Sheila looked at, ``\textit{What part of the speaking is instruction? What part of the speaking would be giving directions?}'' (RS1). When Bonnie used the Turn-Taking visualization, she noticed, ``\textit{I was talking for this long and there's no student responses happening yet. And I'm scrolling down...and I go `Okay, something occurred here. Oh, what kind of questions was it or was something sticking out?'}'' (RS1). The Turn-Taking and Transcript visualizations together contextualized the quantification seen in the Talk Ratio to give further understanding about class discussions.

In post-interviews, teachers  had generally positive feedback about the Turn-Taking and Transcript visualizations. They confirmed the value of the Transcript for recalling class events. As Tom said, ``\textit{That's what I find shocking is when I would read [the Transcript], I would immediately go ‘Oh! Now I know exactly where I was,’ I would just come back very quickly.}'' Kate felt the Transcript visualization was, ``\textit{the most effective or most significant for me because it really helps me understand the way I was presenting the lesson and how much I was allowing for the students to participate...instead of how I thought I was doing, which aren't always the same thing.}'' Teachers mentioned that the Turn-Taking visualization gave quick insights into the rhythm of discussion. Tom mentioned, ``\textit{It was on [the Turn-taking] that you could click on [the graph] and that will take you to what was being said at that period of time. And I thought that was useful. It made sense to me [and] is easy for me to understand my lesson and access what I wanted to look at.}'' Sheila stated how the Turn-Taking gave visibility to her discussion facilitation, ``\textit{Just by scanning through [the Turn-Taking], I think you can tell [if] you're giving your students time to actually talk, if you're preparing them well enough in order to discuss on their own, or giving them the opportunity to talk to each other to respond individually.}'' Kate also specifically mentioned the connection between the Turn-Taking and Transcript visualizations in the miniview, ``\textit{I think this [miniview] on the left is very helpful and then I do think that seeing the dialogue as well is really helpful to see who's actually engaged...so I think this page together is very helpful.}'' However, Jeff was critical of the two visualizations, stating, ``\textit{I probably wouldn't look at what [students] said...What they say really isn't as important as how much they say.}'' He noted instead, ``\textit{To me [the Turn-Taking] needs to be broken up by students so just one accumulated bar for student one, [another] the bar for student two because I want to see which student is talking how much.}''

\subsection{Shifting Professional Vision}
Where contextualization refers to how teachers viewed and reflected on past data for understanding, we observed how teachers shifted in their professional vision in which they examined their data with an eye towards future actions.

\subsubsection{Early Noticings: Quantification of Discussion}
Related to our findings about quantification, teachers often focused on the amount of student talk in early reflection sessions, which may reflect their level of noticing \cite{van2021expanding}. This led teachers to evaluate their data. Kate evaluated her use of talk codes, ``\textit{The student breakout group shows...Making Reasoning Explicit and Building on Ideas. Those parts are definitely good}'' (RS1). Jeff evaluated the discussion cadence, ``\textit{There was a good portion where it seemed like it was kind of back and forth. But maybe not fully reaching that goal, something that I still need to continue to be aware of}'' (RS2). Bonnie noted specific wording, ``\textit{I think it was the wording in the question, and I know how I would do it differently next year for sure...I was hoping to get at least a few questions kind of like, you know, touched upon.}'' However, evaluation of data could also lead to negative feelings. In her second reflection session, Sheila said, ``\textit{[I] still...feel so inadequate. It's really sad because...it's really hard to get them engaged}'' (RS2). Though we observed this in one teacher, Sheila only had two reflection sessions so we were unable to see how or if these were resolved. 

With an emphasis in noticing quantitative aspects of talk, teachers set quantitative goals to increase student talk in general in early reflection sessions. Bonnie stated her goal to ``\textit{have less teacher talk and more student interaction because that's where more learning takes place}'' (RS1). Jeff set goals around how many students spoke, ``\textit{I'd say that the types of goals we want to set...would maybe be the number of people that interact, trying to reach a certain percentage of people interacting in conversations}'' (RS1). These goals reflected how teachers interpreted their data at their stage of professional learning.

\subsubsection{Shifting Expectations of Data}
Over multiple reflection sessions in engaging with the tool, teachers' professional vision shifted in their expectations of data. Early on, Bonnie did not know what to expect in her Turn-Taking visualization, ``\textit{I didn't have an expectation actually...No expectation, I didn't know what I was gonna see}'' (RS2). By her last reflection, she knew what to look for in her data, ``\textit{This is the kind of stuff I wait for, is what do they understand about the new concepts we're getting into?}'' (RS5). Kate began to notice how her actions impacted discussion, ``\textit{I was happy to see the Build on Ideas because...I had been purposeful about wanting to get them to remember what we'd done the week before. So it was nice to see that that was actually captured in the data}'' (RS4). Notably, Kate spoke about her expectations and strategies in terms of the talk codes. Jeff also showed this shift. In an early reflection, he stated, ``\textit{So they're making connections. That's what I want}'' (RS2). During his last reflection session, he was more specific, ``\textit{A lot of times before the students would say...very short answers, but now we're getting some multiple sentence ideas...it seems like we're getting more complete thoughts out of each student this way...When I had put them in small groups before, it didn't look like this}'' (RS6). Over time, we saw how teachers had more directed noticings and expectations in their data.

We also observed two teachers set goals beyond quantitative talk goals. Bonnie set goals related to types of talk, ``\textit{I wanted to put that as a goal for discussion, that they're building on each other's ideas or they're using some of it to reformulate their own}'' (RS3). Kate deliberately shifted away from looking at the quantity of talk, ``\textit{We had mentioned that we weren't really gonna focus on the Talk Ratio}'' (RS3). She set a goal for more student evaluation, ``\textit{I would like to get to the point where I could have students interacting with each other and have them evaluate each other}'' (RS3). However, Jeff preferred the Talk Ratio visualization through all his reflection sessions, ``\textit{Mostly what tells me [that I'm moving forward] is where this line is in the middle...The Talk Ratio is probably still...the most useful thing for me.}'' All teachers changed in their expectations of data, but quantification of talk sometimes super-ceded focusing on other aspects of discussion. 


\subsection{Adoptability: Factors and Barriers for Adoption of Conversation Support Technology}
In post-interviews, teachers addressed multiple aspects of conversation support tools and their implementation that would lead to or hinder adoption and continued usage of these sorts of technology.

\subsubsection{Personalization, Persistence, and Regularity of Data}
Teachers expressed frustration that prior PD experiences did not seem applicable to their own classrooms and contexts. Sheila noted, ``\textit{[We] will get curriculum from people who have written the curriculum for our grade level...and we're like who are the students that these people are writing this for? Definitely not mine.}'' Tom also said, ``\textit{Teachers would be forced to change, but [the PD professionals] don't know why a certain teacher is effective or why they're successful with their students.}'' In contrast, teachers appreciated the personalized PD provided in our tool. Kate stated, ``\textit{This is the only PD that's been specific to me and my behavior in the classroom and my presentation of lessons...this was like getting into the nitty gritty of how I actually am in the classroom and no PD has ever even come close to that even when I get evaluated every two years.}'' Sheila thought, ``\textit{Any data is helpful to see what I'm doing great and what I'm doing that I should be better at…This would be something that we have clear, concrete data as to…what did we do, are we meeting reaching our own personal goals that we're actually putting forth?}'' Teachers felt the data within the tool provided personalized feedback where they could see their growth whereas generalized seminars did not. 

Teachers also expressed frustration at the lack of follow-up and accountability from prior PD seminars and workshops. Sheila said, ``\textit{PD that we have are all `here you go, implement'... Typically it's a one and done...and there is not any real accountability as to whether or not you're even doing anything.}'' Kate concurred, ``\textit{Usually PD stops at the idea phase like 'oh here's a great idea to engage your students, now go do it.'}'' Jeff also mentioned that the lack of follow through, ``\textit{Our big complaint as teachers has been we never see [the PD] again, and we never analyze how we do things are changed}.'' These statements from teachers align with findings that PD is often neither personalized nor persistent \cite{jayaram2012breaking,popova2022teacher,ball1999developing,sherin2008professional}. In contrast, teachers appreciated the persistence of multiple reflection sessions. As Tom said, ``\textit{You guys have been more persistent…Because you keep on coming back and you keep on reminding me, `Okay, this is what we did last time.}''' Jeff said this form of PD was ``\textit{more about the progress in the journey...This process has been I think much more useful being long-term rather than these one and done things that most districts do.}'' Tom valued the accountability of the reflections, ``\textit{What I'm getting out of this is it was requiring me to reflect, which teachers should do anyways…I know some percentage do, but I don't think it's the whole group.}'' Sheila added that separation from the district was an important factor, ``\textit{This is not intrusive. [It’s] low risk, given that…it's helping me stay accountable to me because I'm also accountable to you [as researchers], but you're not accountable to my district.}'' However, challenges in data collection and coding led to large gaps between classes and reflection sessions. After long periods of time, the Transcript may be less effective in recalling class events. As Tom said, ``\textit{There would be this delay in terms of...[the reflection session] and when the lesson was. So...I was asked to look at [the data] ahead of time, which I would do, but that would be an extra step...I'd be asked these questions and at the same time I'd be trying...to remember what was said.}'' We found that persistence of reflection is important if done with somewhat consistent regularity. These findings suggest a need to restructure teacher PD to enable greater personalization and shift away from the ``one-and-done'' nature of current PD practices.

\subsubsection{Learning Curve of Technology and Talk Codes}
Teachers expressed a learning curve in both using our prototype as well as understanding the talk codes. Some of the barriers came from a general resistance to technology. As Tom stated, ``\textit{Teachers need to take advantage of the technology that's available and they're not. A lot of teachers are technology phobic and you know they're being forced to learn how to use programs and stuff.}'' Bonnie stated a learning curve specifically in using our prototype, ``\textit{First of all, just navigating [the app]...that in itself is a learning curve. There's two things happening here. One is how was the lesson? And the other is how is the app?...So there's several moving parts here}.'' Tom also talked about challenges in learning to understand and interpret the data visualizations and talk codes, ``\textit{So brown is Invite...I wasn't quite sure how it was being decided that the snippets that I was reading, um, how that matched with invite. Because I would've thought that all of mine would've been brown, because I'm always asking the students, 'okay, what do you think?'}'' (RS5). Jeff also expressed confusion about the talk categories, ``\textit{I'm sitting here looking back at the [legend] and saying, 'okay, that one's Connect. And then over here, Evaluation.' I'm thinking, how's Evaluation different from Reasoning?} (RS6). This is a limitation of developing a shared understanding of discussion theory and practice between researchers and teachers.

\subsubsection{Granularity and Types of Data Presented}
Related to the learning curve of interpreting data within the tool was the granularity of data presented. Jeff mentioned the data in the Turn-Taking and Transcript visualizations was too fine-grained, ``\textit{I thought that the very basic Talk Ratio is helpful, but a lot of the other things...were too much information and not really useful...The granular level of detail...[for] a teacher using a daily or weekly tool, it's just too much information [and] too time consuming.}'' Bonnie found it challenging to navigate through specific talk codes, ``\textit{This Other Teacher Talk is really interesting...I'd have to go through each [excerpt] to know}'' (RS2). She also felt that talk data did not capture the spectrum of student learning, ``\textit{Just because they're not saying [anything] doesn't mean they're not writing an elaborate report or talking about it with each other}'' (RS2). Jeff suggested measures related to individual student talk, ``\textit{I'd like to see what's the total [talk] for student one? What's the total for student two?}'' (RS3). Tom wanted to see silence or wait time reflected, ``\textit{If I didn't say anything for 15 minutes, there would just be no timestamps. There's no space indicating that there was 15 minutes of quiet time}'' (RS3). These different forms of data could help teachers reflect more deeply on their discussion.

\subsubsection{Generalization of Discourse Visualization}
Beyond our reflection sessions, teachers mentioned other situations where personalized discourse analysis would be useful. Kate suggested collaboration with other teachers, ``\textit{I would love to have...a colleague cohort working with [data], but I would use it, even if I was just on my own. I think it's that helpful and valuable in terms of improving the quality of teaching.}'' Sheila thought discussion data could be useful for new teachers, ``\textit{I would think for any new teacher...unless someone’s watching you and giving you feedback, you don’t know what you’re doing.}'' Teachers also mentioned discourse analysis for other settings. Tom stated how discussion data would be useful in supervision meetings, ``\textit{I was supervising maybe four people and...you spend a lot of time thinking about...what they're doing, what you want them to do, and how to...achieve that goal. So that's an opportunity where...you would be interested in recalling specific conversations and you know what was said.}''  Kate also thought this data could be useful for staff meetings, ``\textit{We have staff development dates coming up, and I would love to have this kind of data for staff development because I don't think the facilitator, who is also the principal, really understands he talks 95\% of the time…I would love to see it in any kind of a group meeting}.'' Jeff similarly thought, ``\textit{I think it might be interesting if they looked at...this sort of thing but looking at professional development, and how teachers interact with instructors and administrators...It's kind of funny how a lot of the [PD facilitators do] exactly the opposite of what they tell us to do.}'' Teachers found discourse analysis useful both for their own insights as well as other situations where interactions are ephemeral and subject to personal perceptions.

%% file: sections/7_Discussion.tex
\section{Discussion \& Future Work}
In answering our research questions, \textit{1) How did teachers use features of the app in their reflections over time? and 2) What are the factors and barriers of adoption of discourse visualization tools?}, we found how teachers used data visualizations in our tool to understand their discussion practices towards future actions. The themes of quantification, contextualization, shifting professional vision, and adoptability cut across these questions. In this section, we discuss lessons learned from our longitudinal deployment and design implications that might generalize to other professions.

\subsection{Design Implications}

\subsubsection{Resolving Dissonance}
The Talk Ratio visualization was designed to answer the question of how much teachers talked versus students. Several systems like the M-Powering Teachers tool \cite{demszky2023can} and TeachFX \cite{ford2021student}, include a similar visualization that quantifies discussion. In answering our first research question, we found that a visualization of talk ratio or talk percentage can be effective at sparking dissonance when the data does not match expectations. As dissonance can cause varying degrees of emotional reaction, it can lead to a desire for resolution \cite{cancino2020general}. We found that over time, teachers in our study sought to resolve dissonance through seeking understanding in the Turn-Taking and Transcript visualizations. This is in line with prior findings that dissonance can motivate self-reflection and understanding \cite{boekaerts2005self,gomoll2021zooming}. However, dissonance may cause negative emotions that may not be resolved in a single reflection session. Sheila, who only had two reflection sessions, noted inadequacies after seeing her Talk Ratio data. In addition, dissonance in quantification of data could emphasize \textit{how much} talk occurred rather than \textit{what types} of talk occurred. It may be that the percentage in the Talk Ratio could lead to inherent evaluation about performance, whether positive or negative. Prior work in personal informatics systems also finds that users may be demotivated if their data shows they did not reach their goals \cite{gulotta2016fostering}. One possible direction to mitigate these reactions is to include positive feedback alongside the quantitative data, similar to the positive prompts the M-Powering Teachers tool provides \cite{demszky2023can}. Another direction is margin-based design, which includes a range in which users could achieve their goals. Jung et al \cite{jung2021good} found that setting goals within a margin allowed users to evaluate their behaviors as ``good enough'' rather than a failure. These directions could help to reduce negative emotions associated with dissonance and lead to productive reflection.

\subsubsection{Scaffolding Attention to Relevant Data}
Teachers generally found the Turn-Taking and Transcript visualizations useful for adding context to the Talk Ratio data and resolving potential dissonance. We noticed that teachers changed in their expectations of this data over time as they viewed these visualizations, representing a shift in their professional vision. However, some teachers found the data too granular. As a result, their early reflections and goals centered around the Talk Ratio visualization and quantification of talk. One potential reason is that the Talk Ratio was the first visualization teachers saw and was separated from the more closely-integrated Turn-Taking and Transcript visualizations, making it more challenging to connect patterns between the three visualizations. A potential implication is creating layered visualizations that tell a narrative about teachers' discussion data through storytelling elements that allow for both coarse-grained exploration and fine-grained explanation in data \cite{echeverria2018exploratory,martinez2020data}. Data annotations that extract relevant points in the data can also scaffold sense-making in complex visualizations \cite{hullman2013contextifier,echeverria2018exploratory}. In our own work, we are currently building annotations into the data visualizations for teachers to better connect their reflection notes to specific points in the data.  

A larger discussion is what ``relevant'' data means. As part of this research-practice partnership, our goal was to increase teachers' usage of academically productive talk and designed our tool around this goal. Some teachers found the talk codes difficult to actually apply in their classroom discussions and instead mentioned aspects of discussion they found more relevant, such as how much individual students spoke or wait time. While researchers explained talk codes to teachers during reflection sessions, these codes and definitions were not co-constructed or designed with teachers as they were informed by prior research. A mismatch between what measures researchers value versus the views of teachers (and students) may hinder adoption and acceptance of technology in practice \cite{ngoon2023,niemi2002active}. Developing a shared understanding of discourse terminology and its meanings is an ongoing challenge in research-practice translation \cite{tripathi2022designing}. Co-designing definitions or terminology that fit within teachers' understanding of talk and discourse may improve how teachers interpret their discourse data. Another implication is to progressively reveal parts of the data that foreground and background elements of the visualizations according to teachers' own professional vision. We found how teachers' shifted in their expectations of data and the goals they set from quantitative to characteristics of the discussion. Scaffolding professional vision through hints, notifications, or other guidance for what and when to analyze classroom data could proivde adaptive support to teachers in examining their discussion data \cite{michalsky2020preservice}. This is an important area that requires close conversation and collaboration between teachers and research and design teams. 

\subsubsection{Discourse Analysis Tools for Professions Beyond Teaching}
Beyond teaching, many other professions rely on quality interactions between professionals and those they serve. These professions include those in healthcare, mental health, coaching, and customer service. Prior work on automated systems in the professions often focuses on productivity such as systems in algorithmic management \cite{bakewell2018everything,lee2015working} and personal informatics systems to track workers' time usage \cite{epstein2016taking}. Conversation analysis using instrumented sensors (such as microphones) to improve professional interactions is a growing area \cite{hirsch2018s,liu2016web,liu2016improving,gerritsen2018towards}. For instance, psychotherapists valued the automated feedback about how they converse with patients the CORE-MI system provided \cite{hirsch2018s}. Liu et al \cite{liu2016web,liu2016improving} found that a system that visualizes non-verbal behaviors improved medical students' awareness of these behaviors in doctor-patient rapport. These professions, like teaching, have a set of established best practices and also lack ongoing, personalized feedback and PD. Teachers themselves mentioned other situations where conversation support could help generate dissonance towards behavioral change and in recalling specifics about interactions. Our themes from this work could apply to professions for which discourse and interaction are key components of professional success, but are not easily quantified or evaluated. Future work could explore what these types of interfaces could look like in other professions for broader professional learning.

\subsection{Challenges with Classroom Studies and Implications for Scaling Conversation Support}

\subsubsection{Challenges of Authentic Classroom Studies and Alternative Implementation Models}

Our model of reflection consisted of collaborative reflection sessions where teachers discuss their discussion data in depth with a PD researcher. This model is line with instructional coaching and personalized teaching consultations where PD experts observe teachers' progress over time and provide feedback \cite{desimone2017instructional,jayaram2012breaking}. Our findings are closely tied to the model in which they were situated, collaborative reflection sessions with a PD researcher that covered one class session in-depth. While teachers appreciated the personalized and persistent PD provided, constraints in scheduling, the COVID-19 pandemic, and data processing and coding time meant that reflection sessions did not happen at consistent intervals. This irregularity may have impacted teachers' reflection of the data though we observed changes in teachers' noticings and expectations of their data even with these challenges. This provides promise for different implementation models of reflection using discourse visualization tools. For example, because some schools may not have the resources to provide teachers with regular one-on-one PD, teachers mentioned possibly creating communities of practice between peers or mentorship communities between experienced and novice peers \cite{patton2017teacher}. A structural model might be setting time aside specifically for teaching reflection. School administrations could create these structures for teachers to record their own classes and take the time to reflect on their data themselves or with peers. This could provide teachers with agency in their reflection and incorporation of classroom technologies \cite{varanasi_reconfigure}.

\subsubsection{Scaling Conversation Support with AI}
For self- or peer-regulated reflection structures to occur, automation of transcribing discussions and categorizing talk is necessary. In this current iteration, discourse was human-coded for ground truth accuracy, which is a labor-intensive and time-consuming process. Several automated models can classify discourse with accuracy on par with that of humans and can scale teacher feedback on discussion \cite{d2015multimodal,jensen2020toward,jensen2021deep,schlotterbeck2021classroom}. We are currently working on automated models that can reliably classify the discourse categorized in our tool. However, even with human-coding, there are challenges with inter-rater reliability and agreement in talk categories \cite{clarke2023developing}, and people in general may not trust AI judgments due to lack of transparency \cite{nazaretsky2022instrument,zhang2020effect}. We found that teachers expressed confusion in how the talk codes were categorized even with human coding, which may impact trust and how teachers might perceive any feedback provided from an AI system. A mixed-initiative approach in which users can evaluate and refine automated outputs could create a collaborative feedback loop \cite{deterding2017mixed}. Increasing transparency in how AI judgments are made could improve trust through explanations alongside confidence scores to explain where models are potentially less accurate. Our own future work is exploring designs with code correction and confidence scores. Other work could expand on how teachers perceive the accuracy and usefulness of AI feedback.

\subsubsection{Privacy Implications}
We focused on audio recordings of discussion for this work, which has limitations in capturing the full spectrum of learning behaviors. Multimodal data beyond audio (such as video or wearables) could capture both verbal and nonverbal behaviors \cite{d2015multimodal,martinez2020data,martinez2020teacher}. However, these modalities of data (including audio data) are part of a significant conversation around privacy concerns, particularly with the involvement of K-12 students and parental consent. State laws and district-level policies dictated how we collected classroom data and what types of data we could collect. We had multiple discussions with school leadership and our institution's IRB to ensure informed consent from participants as well as compliance to laws and policies. As laws vary depending on location, navigating these restrictions could be a challenge in automated discourse support. In addition, sharing of data is an important consideration. In our study, teachers mentioned that the separation between the research team and administration was the reason why they were comfortable sharing their data with us. They may have felt differently if administration was more involved in the use and analysis of their data. Since schools, districts, or PD organizations are likely to be the stakeholders who purchase and implement these types of conversation support tools, future research on guidelines around the collection, use, and sharing of data are necessary to move this field forward.

\subsection{Limitations}

There are several limitations to this work. The COVID-19 pandemic caused major disruptions to our data collection workflow. As classes were shifted online, data collection paused for a portion of the 2020-2021 school year and resumed during online teaching. While online teaching did make audio recording easier due to built-in recording functions (\textit{e.g.} the Record function in Zoom), it also led to significant differences in classroom discussion behaviors. Teachers reported that their students engaged far less in discussion than in in-person classes. As a result, there were large discrepancies in teacher versus student talk for courses recorded online, which may have impacted how teachers reflected on their data towards their discussion goals during this period. In addition, the regularity with which reflection sessions could be scheduled was impacted by the teachers' schedules, time constraints of talk coding, and the pandemic. This meant that reflection sessions sometimes occurred months after data was collected, which could affect teachers' memory of the specific class and how they might take action towards their discussion goals. Student behavioral and learning outcomes in class over time from reflections were not in the scope of this paper, but these would likely influence how teachers engaged with their data. Lastly, our sample size of teachers was small with one of the 5 teachers in our study being unable to participate in interviews. However, the reflection sessions provide a rich longitudinal data set for understanding teachers' reflections on their discussion data in ClassInSight. 

\section{Conclusion}
In this paper, we present ClassInSight, a tool that visualizes conversational data to provide teachers with personalized PD in their classroom discussions. From 22 reflection sessions and interviews with 4 middle and high school teachers, we found that themes of \textit{quantification, contextualization, shifting professional vision},  and \textit{adoptability} affected how teachers interacted with their data and the affordances of the tool. Over time, their professional vision developed from simply evaluating their data to generating expectations and questions about their data and the context of discussions. We contribute an understanding of how design can impact teacher reflection over the course of a longitudinal deployment. We discuss the design implications of these themes for conversational support for professions where conversation is a critical aspect for professional success.

\begin{acks}
We sincerely thank our teacher participants for their time and participation within this longitudinal study. We also thank Daniel Noh, Lesley Yan, Sara Bai, Gabriella Howse, and Lucia Fang for their design and data analysis efforts. This work is supported by the McDonnell Foundation \#1822813 and a NSF STEM Education Postdoctoral Research Fellowship \#2222530.
\end{acks}